\documentclass[aip,rsi,reprint,floatfix]{revtex4-1}  
\usepackage{graphicx}  
\usepackage{dcolumn}   
\usepackage{bm}        
\usepackage{amssymb}   
 \usepackage{float}
\usepackage{amsfonts}
\usepackage{gensymb}
\usepackage{amsmath}
\usepackage{upgreek}
 \usepackage{textcomp}
 \usepackage{hyperref}

\hypersetup{
    colorlinks=true,
    linkcolor=blue,
    filecolor=magenta,      
    urlcolor=magenta,
}
 \usepackage{mathtools}

\DeclarePairedDelimiter\ket{\lvert}{\rangle}
\DeclarePairedDelimiterX\braket[2]{\langle}{\rangle}{#1 \delimsize\vert #2}
\usepackage{physics}
\usepackage{blindtext,tikz}
\usetikzlibrary{calc}
\nocite{*}

\usepackage{listings}
\usepackage{color}

\definecolor{dkgreen}{rgb}{0,0.6,0}
\definecolor{gray}{rgb}{0.5,0.5,0.5}
\definecolor{mauve}{rgb}{0.58,0,0.82}

\begin{document}
\title{CIF2WAN: A Tool to Generate Input Files for Electronic Structure Calculations with Wannier90}

\author{Christopher Sims}
\affiliation{\textit{ Department of Physics, University of Central Florida, Orlando, Florida 32816, USA}}
\thanks{Christophersims@knights.ucf.edu}

\begin{abstract}
The generation of input files for density functional theory (DFT) programs must often be manually done by researchers. If one wishes to produce a maximally localized wannier functions (MLWFs) the calculation consists of several separate files that must be formatted correctly in order for the program to work properly. Many of the inputs are repeated throughout the files and can be easily automated. In this work, a program is presented to generate all of the input files needed to produce wannier functions with Wannier90 starting from open source DFT programs such as Quantum espresso, Abinit, and Siesta. In addition, the input files for WannierTools are also included for those that wish to produce surface green's functions for the generation of surface state bands. The program presented allows for users new to DFT to use the programs with minimal understanding of parameters needed to produce good results, in addition, this program allows for researchers who are advanced DFT users to utilize this program for high throughput wannier calculations.

\end{abstract}
\maketitle
\section{Program Summary}

\noindent \textit{Program title}: CIF2WAN \\
\textit{Licensing provisions}: GNU General Public Licence 3.0\\
\textit{Program obtainable from}:
\textit{Programming language}: Python\\
\textit{Has the code been vectorised or parallelized?}: no\\
\textit{Computer}: Any computer that can run Python v3.6+\\
\textit{Operating system}: Any operating system that can run Python v3.6+\\
\textit{external libraries}: numpy, pymatgen (w/ registration), glob, shutil, csv\\
\textit{Running time}: less than a minute (DFT runs are separate) \\

\section{Introduction}
Density functional theory (DFT) is a powerful tool that is widely used for calculations in solid-state physics such as electronic structure predictions, lattice relaxation calculations, magnetism, etc. Until the late 1990's, DFT was considered to be too computationally expensive for such calculations\cite{Kohn1965,Jones2015}. However, with increasing technology and computational methods DFT is nor considered to be accurate enough for quantum chemistry. There are a plethora of DFT packages for quantum chemistry such as VASP\cite{VASP-Kresse2-PhysRevB.48.13115}\cite{pbegga}, ABINIT\cite{Gonze2009,Gonze2002}, Quantum ESPRESSO\cite{Giannozzi2017,Giannozzi2009}, SIESTA\cite{Soler2002}, and WIEN2K\cite{Blaha2020} being the most popular. For real space electronic structure calculations, It is necessary to extract the real space maximally localized wannier functions (MLFWs)\cite{Marzari1997,Souza2001,Marzari2012} in order to calculate the surface band structure \cite{sancho1985}, perform wannier charge center calculations, etc. There are two main programs that are capable of calculating MLWFs one being Wannier Transport (wanT)\cite{Calzolari2004} and the other, more popular being Wannier90\cite{Pizzi2020}. With these programs on is able to calculate surface band structures utilizing the iterative surface green function matching technique such as the method implemented in Wannier tools\cite{wtools} or chinook\cite{Day2019}. The main issue with DFT, especially if one wishes to conduct high throughput calculation is that one must meticulously format the input files for DFT programs so that there is no error in the calculation. Small errors cannot bypass the error detection built into DFT programs and learning the best methods for input will take some time for a beginner. Rather, having programs that assist in the pre-processing stage are more favorable for \textit{ab initio} DFT theorists. While tools such as CIF2CELL\cite{Bjoerkman2011}, Aiida\cite{Pizzi2016}, pymatgen\cite{Jain2013}, jarvis\cite{Choudhary2017,Choudhary2018,Choudhary2018a,Choudhary2019,Choudhary2019a,Choudhary2020}, etc. exist, non can generate all of the input files needed to generate the wannier tight binding (TB) Hamiltonian utilizing multiple different quantum chemistry programs.

Recently, there has been a rising interest in identifying materials by their surface states. This began with the discovery of a 3D topological insulator which was confirmed by angle-resolved photoemission spectroscopy and density functional theory\cite{Hasan2010,Hsieh2009b,Xia2009a}. Later, nodal-line and Weyl semimetal states have been discovered in real crystal systems such as ZrSiS\cite{Hu2016,Schoop2016,Neupane2016a,Hosen2017} and TaAs\cite{Xu2015,Lv2015,Huang2015}, respectfully. All of these materials are confirmed to be topologically non-trivial via berry phase, wannier charge center, or Wilson loop analysis. All of these methods can be performed with wannier functions. However, in order to properly perform calculations of a large amount of materials that can be topologically non-trivial will require many input files to various DFT programs. The automation of these tasks will make these calculations much simpler to perform.

This work presents CIF2WAN, a python program that can generate the input files in order to perform wannier calculations from \textit{ab initio} calculations. In addition, slurm job handling files which are common in most supercomputer clusters are also generated for the user. The goal of this program is to make DFT calculations that utilize wannier TB Hamiltonians easier to understand for beginners and easier for more advance users of DFT programs.
\section{Methods}
Wannier functions can be seen as a derivation of the bloch function with an associated phase phase factor $e^{-i\mathrm{\textbf{k}}\cdot\mathrm{\textbf{R}}}$ is the real space lattice vector. In addition, one can label each band with its associated real space lattice vector $\mathrm{\textbf{R}}_n$ where $n$ is the band index. With this we can consruct the wannier functions for each band index $n$.\cite{Wannier1937}
\begin{equation}
\ket{\mathrm{\textbf{R}}_n} = \frac{V}{(2\pi)^3}\int_{BZ} d\mathrm{\textbf{k}}e^{-i\mathrm{\textbf{k}}\cdot\mathrm{\textbf{R}}}\ket{\Psi_{n\mathrm{\textbf{k}}}}
\end{equation}

In order to compute wannier functions one must introduce a unitary mixing parameter $U_{mn}^{\textbf{k}}$ to make the Hamiltonian smooth in all of k-space 
\begin{equation}
\ket{\mathrm{\textbf{R}}_n} = \frac{V}{(2\pi)^3}\int_{BZ} d\mathrm{\textbf{k}}e^{-i\mathrm{\textbf{k}}\cdot\mathrm{\textbf{R}}} \sum_{m=1}^J U_{mn}^{\textbf{k}}\ket{\Psi_{n\mathrm{\textbf{k}}}}
\end{equation}
A common way to compute the wannier functions is via the projection method. Starting from a set of trial projections ($J$), These trial functions are projected onto the Bloch manifold.
\begin{equation}
\ket{\psi_{n\mathrm{\textbf{k}}}} = \sum_{m=1}^J\ket{\Psi_{m\mathrm{\textbf{k}}}}\braket{\Psi_{m\mathrm{\textbf{k}}}}{g_n}
\label{trial}
\end{equation}
In order to do this calculation the matrix of inner products must first be computed. $(A_{\mathrm{\textbf{k}}})_{mn} = \braket{\Psi_{m\mathrm{\textbf{K}}}}{g_n}$. By substituting this into equation \ref{trial}, one can construct the trial wannier functions that are related to the real Bloch functions.
\begin{equation}
\ket{\widetilde{\psi}_{n\mathrm{\textbf{k}}}} = \sum_{m=1}^J\ket{\Psi_{m\mathrm{\textbf{k}}}}(S_{\textrm{\textbf{k}}}^{-1/2})_{mn}
\label{Twan}
\end{equation}
Using these equations one can then minimize the localization function $\Omega$
\begin{equation}
\Omega= \sum_n[\mel{\textbf{0}n}{r^2}{\textbf{0}m} - \mel{\textbf{0}n}{\textbf{r}}{\textbf{0}n}^2] = \sum_n[r^2 - \tilde{\textbf{r}}_n^2]
\end{equation}
MLWFs can be derived by minimizing this function for each k point in the lattice with respect to $U_{mn}^{\textbf{k}}$ after the bands have been calculated during the self consistent step of a DFT calculation. This procedure is repeated until $\Delta\Omega$ is sufficient small.\cite{Marzari2012}
\section{Features of CIF2WAN}\label{feat} 
CIF2WAN generates all of the input files needed to start from \textit{ab initio} calculations in ABINIT, Quantum ESPRESSO, etc. to the generation and use of the wannier tight binding Hamiltonian (seedname\_hr.dat) used in wannier tools. The program can load these files utilizing either a CIF file or by directly interfacing with materials project via pymatgen. The output is as follows for the following DFT programs.
\begin{table}[ht]
\caption{Output for Quantum ESPRESSO}
\begin{tabular}{l|l}
\hline
File             & Description\\ \hline \hline
seedname.scf.in             & self consistent input file\\ \hline
seedname.nscf.in             & Non-self consistent input file \\ \hline
seedname.p2w.in        & input file for PW2Wannier90  \\ \hline
seedname.win       & input file for wannier90 \\ \hline
scf.slurm         & scf slurm file \\ \hline
nscf.slurm      & nscf slurm file  \\ \hline
p2w.slurm      & p2w slurm file \\ \hline
wann.slurm     & wann slurm file \\ \hline
cleandft.sh & cleanup for another DFT run \\ \hline \hline
PP/     & Pseudopotential folder \\ \hline
WT/          & wanniertools folder  \\ \hline
WT/wt.in  & wanniertools input file\\ \hline
WT/wt.slurm  & wanniertools slurm input file\\ \hline \hline
\end{tabular}
\label{QE}
\end{table}
\begin{table}[!ht]
\caption{Output for ABINIT}
\begin{tabular}{l|l}
\hline
File             & Description\\ \hline \hline
seedname.in             & ABINIT input file\\ \hline
seedname.files             & file linking for ABINIT\\ \hline
seedname.slurm        & main slurm file  \\ \hline
w90.win       & input file for wannier90 \\ \hline
cleandft.sh & cleanup for another DFT run \\ \hline \hline
PP/     & Pseudopotential folder \\ \hline
WT/          & wanniertools folder  \\ \hline
WT/wt.in  & wanniertools input file\\ \hline
WT/wt.slurm  & wanniertools slurm input file\\ \hline \hline
\end{tabular}
\label{ABI}
\end{table}
\begin{table}[!ht]
\caption{Output for VASP}
\begin{tabular}{l|l}
\hline
File             & Description\\ \hline \hline
PBE/		& Folder for PBE run \\ \hline
HSE06/		& Folder for HSE run \\ \hline
W90/		& Folder for W90 run \\ \hline
KPOINTS            & Global, K points \\ \hline
POSCAR            & Global, crystal information\\ \hline
PBE/INCAR             & INCAR for PBE\\ \hline
HSE06/INCAR             & INCAR for HSE06\\ \hline
W90/INCAR             & INCAR for W90\\ \hline
vsp.slurm        & main slurm file  \\ \hline
wannier90.win       & input file for wannier90 \\ \hline
WT/          & wanniertools folder  \\ \hline
WT/wt.in  & wanniertools input file\\ \hline
WT/wt.slurm  & wanniertools slurm input file\\ \hline \hline
\end{tabular}
\label{VASP}
\end{table}
\begin{table}[!ht]
\caption{Output for SIESTA}
\begin{tabular}{l|l}
\hline
File             & Description\\ \hline \hline
seedname.fdf		& main input file\\ \hline
*.psf		& pseudo for atoms\\ \hline
wannier90.win       & input file for wannier90 \\ \hline
cleandft.sh & cleanup for another DFT run \\ \hline \hline
WT/          & wanniertools folder  \\ \hline
WT/wt.in  & wanniertools input file\\ \hline
WT/wt.slurm  & wanniertools slurm input file\\ \hline \hline
\end{tabular}
\label{VASP}
\end{table}
\section{Installation and usage}
In this section, we present how to install and use CIF2WAN.
\subsection{Get CIF2WAN}
\textsc{CIF2WAN} is an open source software package distributed under the GNU General Public license 3.0 (GPL). The code can be downloaded directly from the public code repository: \href{https://github.com/ChristopherSims/CIF2WAN}{https://github.com/ChristopherSims/CIF2WAN}.
\subsection{Installation and running the code}
The code requires no installation, in the future this package may be developed into a python package. The version of python must be 3.6 or higher, in addition the user must install the following python packages: pymatgen, matplotlib, numpy, string, shutil, csv. These programs will likely automatically install in most modern python CDEs.  In addition, one must obtain an APIKEY for usage of pymatgen (only if one uses the PYMATGEN flag), this key can be obtained by registering an account at \href{materialsproject.org}{materialsproject.org}
\subsection{File formats}
File outputs are as described in section \ref{feat}. The input file must be manually edited by the user in v1.0. However, a GUI may be introduced in later version.
\subsection{Input file}
\begin{lstlisting}[language=Python]
#INPUT FILE input.py
ncore = 50 # number of cores for calculation
USE = "PYMATGEN" # PYMATGEN OR CIF
DFT = "ESPRESSO" #ESPRESSO, SIESTA, ABINIT, VASP
APIKEY = "YOURAPIKEY" #APIKEY for mat
MATERIALID = "mp-22866" # imput for pymatgen
SEEDNAME = "HB" #yours or default
SOC = True # TRUE or FALSE
KMESH = [8,8,8] # KMESH for nscf/wan
NUMBANDS = 120 # Number of bands
MAGNETISM = False
\end{lstlisting}
\section{Wannier90 Convergence} 
Although all of the bands and energies are converged in the scf and nscf cycles there is the main issue of convergence with MLWFs that will cause most calculations to appear to not converge. A small error in the wannier minimization can lead to spurious bands that will lead to a non-real Fermi surface when conducting calculations with the results of that wannier90 run. There are many ways to correct this, such as iterating over many steps ($>$50,000), defining a good guess for the initial projection, changing the frozen window for disentanglement, etc. However, the main issue is two three fold, the initial projections (guess) must be manually defined or be made random, there should be an internal program that makes good projection guesses from the known valence states and the result of the nscf run. Secondly, the steepest decent algorithm implemented in wannier90 v3.1 is not the best algorithm in order to get the proper minimization of the wanner centers. In fact, this method often gets stuck in local minimum during the calculation. Several better alternatives to the algorithm have been proposed but have not been implemented in the main wannier90 code\cite{Mustafa2015,Damle2018,Cornean2019}. Finally, many of the interface codes to wannier90 were not written by the original developers of the respective DFT functions, this means that they could have many bugs or not correctly calculate the .mmn and .amn files needed for wannier90. This creates of problem of there being errors in wannier90 runs that must be corrected manually by users or are so fatal that wannier90 cannot parse to input file correctly at all. detracting from the possibility of \''high throughput'' wannerization runs.
\section{Examples} 
In this section we present calculations utilizing the files that are generated by the programs. The only changes are made to the wannier90 input files in order to converge the MLWFs
\subsection{3D topological insulator Bi$_2$Se$_3$}
\begin{figure}[!ht]
 \begin{center}
 \includegraphics[width=0.5\textwidth]{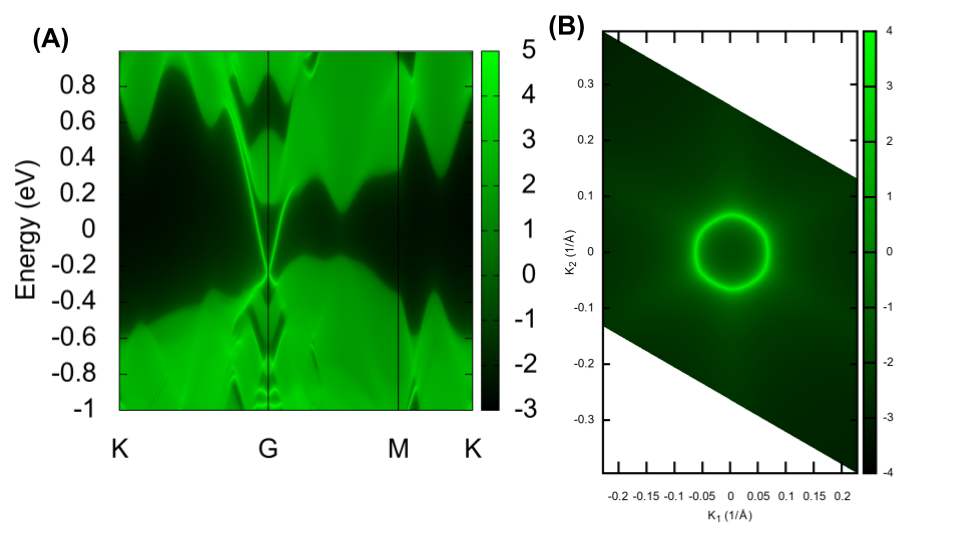}
\end{center}
\caption{\textbf{Bi$_2$Se$_3$}: (A) Surface states of the (001) surface (B) Fermi surface of Bi$_2$Se$_3$}
\label{bs} 
\end{figure}

 Bi$_2$Se$_3$ has been discovered to be a 3D topological insulator\cite{Hsieh2009b,Xia2009a} with non-trivial surface states and an insulating bulk. Here we show a calculation utilizing quantum ESPRESSO with spin orbit coupling (SOC) on. Here we see the large Dirac cone at the $\Gamma$-point [Fig \ref{bs}(A)]. In addition, we show the Fermi surface of the Dirac cone inside the band gap showing the topologically nontrivial surface states[Fig \ref{bs}(B)].

\subsection{Nodal line semimetal HfP$_2$}
\begin{figure}[!h]
 \begin{center}
 \includegraphics[width=0.5\textwidth]{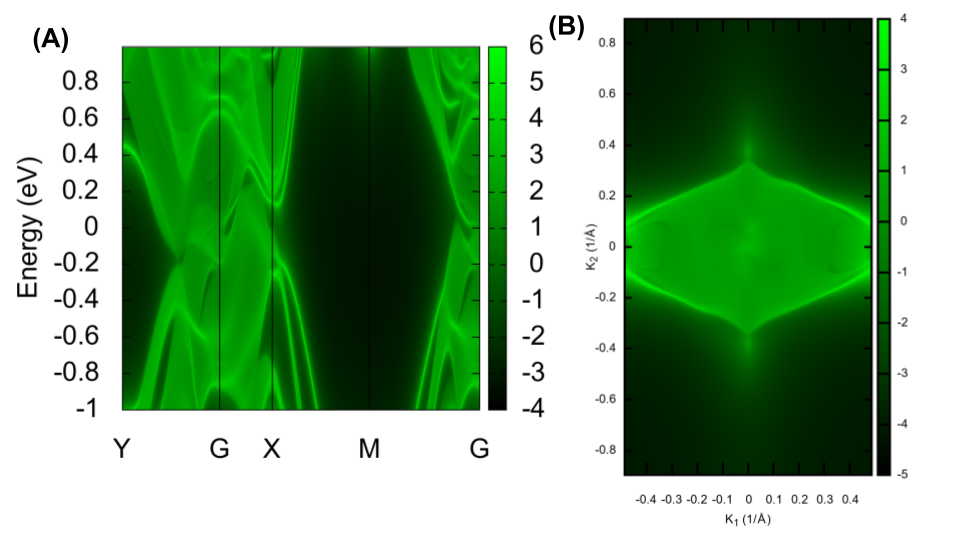}
\end{center}
\caption{\textbf{HfP$_2$}: (A) Surface states of the (001) surface (B) Fermi surface of HfP$_2$}
\label{hp}
\end{figure}

HfP$_2$ is a nodal line semimetal\cite{Sims2019} where the non-trivial state lies above the Fermi level in the presence of negligible spin-orbit coupling. although the nodal-line is above the Fermi surface non-trivial surface states that originate from the nodal line still exist throughout the band structure. The nodal-line is located about 0.2 eV above the Fermi level along the $\Gamma$-$\mathrm{X}$ line [Fig \ref{hp}(A)]. We also show the Fermi surface of HfP$_2$ [Fig \ref{hp}(B)].

\subsection{Weyl Semimetal TaAs}
TaAs is the first material to be discovered to have a Weyl semimetal state. Weyl fermions materialize in condensed matter systems as chiral edge modes which are connected by Fermi arcs.

\begin{figure}[!h]
 \begin{center}
 \includegraphics[width=0.5\textwidth]{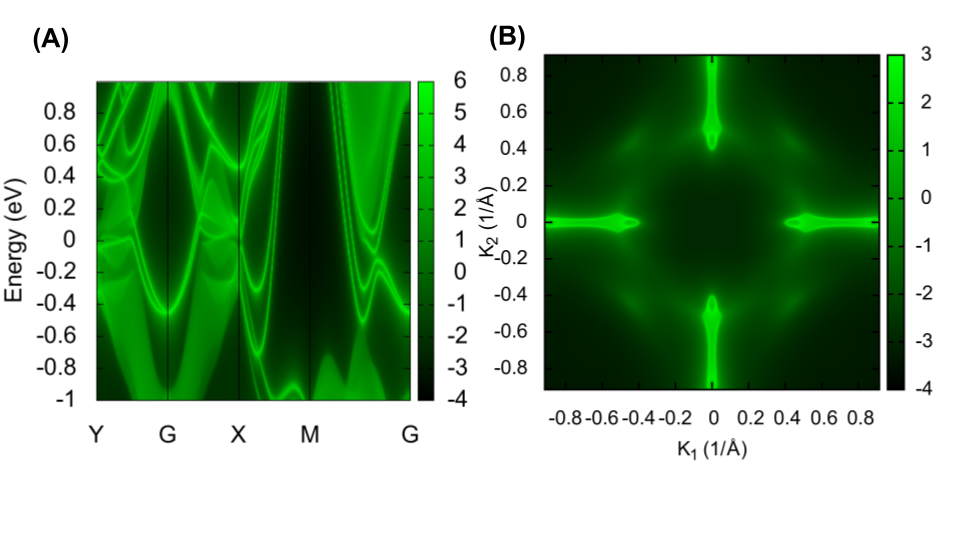}
\end{center}
\caption{\textbf{TaAs}: (A) Surface states of the (001) surface (B) Fermi surface of TaAs}
\label{TA}
\end{figure}


\vspace{-12pt}
\section{Conclusion}

In conclusion we have develop a program that can generate the input files needed to conduct calculations all the way for generating the wannier TB Hamiltonian and utilizing the result in wanniertools. Our program is designed to be easy to use at the expense of complexity and being able to input several settings into DFT programs. Currently, CIF2WAN can interface with quantum ESPRESSO, ABINIT, VASP, and SIESTA, with the possibility of other quantum chemistry programs being added. In the future, we intend to add a graphical user interface once the program has been well tested.

\section{acknowledgments}
The authors acknowledge the University of Central Florida Advanced Research Computing Center for providing computational resources and support that have contributed to results reported herein. URL: \href{https://arcc.ist.ucf.edu}{https://arcc.ist.ucf.edu}.

Correspondence should be addressed to C.S (Email: Christophersims@knights.ucf.edu).
%

\section{Appendix: $\mathrm{Bi}_2\mathrm{Se}_3$}
\subsection{BS.scf.in} 
\begin{lstlisting}
&CONTROL
calculation ='scf'
prefix = 'BS'
outdir = 'bin/'
pseudo_dir = 'PP/'
verbosity='high'
/

&system
ibrav = 0
nat = 5
ntyp = 2
ecutwfc =60 !Ryberg
ecutrho =500
occupations = 'smearing'
smearing = 'gaussian'
degauss = 0.01

!SOC
noncolin = .TRUE.
lspinorb = .TRUE.
starting_magnetization(1) = 0
starting_magnetization(2) = 0


/

&ELECTRONS
conv_thr = 1.0d-7
mixing_beta = 0.495
mixing_mode = 'TF'
diagonalization= 'david'
adaptive_thr=.true.
/

ATOMIC_SPECIES
Bi 208.9804 Bi.rel-pbesol-dn-kjpaw_psl.1.0.0.UPF
Se 78.96 Se.rel-pbesol-dn-kjpaw_psl.1.0.0.UPF


ATOMIC_POSITIONS (crystal)
 Bi 0.3990    0.3990    0.6970
 Bi 0.6010    0.6010    0.3030
 Se 0.0000    0.0000    0.5000
 Se 0.2060    0.2060    0.1180
 Se 0.7940    0.7940    0.8820


CELL_PARAMETERS (angstrom)
-2.069  -3.583614  0.000000     ! crystal lattice information
 2.069  -3.583614  0.000000
 0.000   2.389075  9.546667


K_POINTS (automatic)
8 8 8 1 1 1

\end{lstlisting}
\subsection{BS.nscf.in} 
\begin{lstlisting}
&CONTROL
calculation ='nscf'
prefix = 'BS'
outdir = 'bin/'
pseudo_dir = 'PP/'
verbosity='high'
/

&system
ibrav = 0
nat = 5
ntyp = 2
ecutwfc =60 !Ryberg
ecutrho =500
occupations = 'smearing'
smearing = 'gaussian'
degauss = 0.01

nosym = .true.
nbnd = 200
!SOC
noncolin = .TRUE.
lspinorb = .TRUE.
starting_magnetization(1) = 0
starting_magnetization(2) = 0


/

&ELECTRONS
conv_thr = 1.0d-7
mixing_beta = 0.495
mixing_mode = 'TF'
diagonalization= 'david'
adaptive_thr=.true.
/

ATOMIC_SPECIES
Bi 208.9804 Bi.rel-pbesol-dn-kjpaw_psl.1.0.0.UPF
Se 78.96 Se.rel-pbesol-dn-kjpaw_psl.1.0.0.UPF


ATOMIC_POSITIONS (crystal)
 Bi 0.3990    0.3990    0.6970
 Bi 0.6010    0.6010    0.3030
 Se 0.0000    0.0000    0.5000
 Se 0.2060    0.2060    0.1180
 Se 0.7940    0.7940    0.8820


CELL_PARAMETERS (angstrom)
-2.069  -3.583614  0.000000     ! crystal lattice information
 2.069  -3.583614  0.000000
 0.000   2.389075  9.546667


K_POINTS (crystal)
64
  0.00000000   0.00000000   0.00000000 1
  0.00000000   0.00000000   0.25000000 1
  0.00000000   0.00000000   0.50000000 1
  0.00000000   0.00000000   0.75000000 1
  0.00000000   0.25000000   0.00000000 1
  0.00000000   0.25000000   0.25000000 1
  0.00000000   0.25000000   0.50000000 1
  0.00000000   0.25000000   0.75000000 1
  0.00000000   0.50000000   0.00000000 1
  0.00000000   0.50000000   0.25000000 1
  0.00000000   0.50000000   0.50000000 1
  0.00000000   0.50000000   0.75000000 1
  0.00000000   0.75000000   0.00000000 1
  0.00000000   0.75000000   0.25000000 1
  0.00000000   0.75000000   0.50000000 1
  0.00000000   0.75000000   0.75000000 1
  0.25000000   0.00000000   0.00000000 1
  0.25000000   0.00000000   0.25000000 1
  0.25000000   0.00000000   0.50000000 1
  0.25000000   0.00000000   0.75000000 1
  0.25000000   0.25000000   0.00000000 1
  0.25000000   0.25000000   0.25000000 1
  0.25000000   0.25000000   0.50000000 1
  0.25000000   0.25000000   0.75000000 1
  0.25000000   0.50000000   0.00000000 1
  0.25000000   0.50000000   0.25000000 1
  0.25000000   0.50000000   0.50000000 1
  0.25000000   0.50000000   0.75000000 1
  0.25000000   0.75000000   0.00000000 1
  0.25000000   0.75000000   0.25000000 1
  0.25000000   0.75000000   0.50000000 1
  0.25000000   0.75000000   0.75000000 1
  0.50000000   0.00000000   0.00000000 1
  0.50000000   0.00000000   0.25000000 1
  0.50000000   0.00000000   0.50000000 1
  0.50000000   0.00000000   0.75000000 1
  0.50000000   0.25000000   0.00000000 1
  0.50000000   0.25000000   0.25000000 1
  0.50000000   0.25000000   0.50000000 1
  0.50000000   0.25000000   0.75000000 1
  0.50000000   0.50000000   0.00000000 1
  0.50000000   0.50000000   0.25000000 1
  0.50000000   0.50000000   0.50000000 1
  0.50000000   0.50000000   0.75000000 1
  0.50000000   0.75000000   0.00000000 1
  0.50000000   0.75000000   0.25000000 1
  0.50000000   0.75000000   0.50000000 1
  0.50000000   0.75000000   0.75000000 1
  0.75000000   0.00000000   0.00000000 1
  0.75000000   0.00000000   0.25000000 1
  0.75000000   0.00000000   0.50000000 1
  0.75000000   0.00000000   0.75000000 1
  0.75000000   0.25000000   0.00000000 1
  0.75000000   0.25000000   0.25000000 1
  0.75000000   0.25000000   0.50000000 1
  0.75000000   0.25000000   0.75000000 1
  0.75000000   0.50000000   0.00000000 1
  0.75000000   0.50000000   0.25000000 1
  0.75000000   0.50000000   0.50000000 1
  0.75000000   0.50000000   0.75000000 1
  0.75000000   0.75000000   0.00000000 1
  0.75000000   0.75000000   0.25000000 1
  0.75000000   0.75000000   0.50000000 1
  0.75000000   0.75000000   0.75000000 1

\end{lstlisting}
\subsection{BS.win} 
\begin{lstlisting}
write_hr = .TRUE.
write_xyz = .TRUE.
!wannier_plot = .TRUE. 
spinors = .TRUE.
num_wann = 50
dis_num_iter=1000
trial_step=50
num_iter=2000

exclude_bands = 1-50, 101-200


begin unit_cell_cart
-2.069  -3.583614  0.000000     ! crystal lattice information
 2.069  -3.583614  0.000000
 0.000   2.389075  9.546667
end unit_cell_cart


begin atoms_frac
 Bi 0.3990    0.3990    0.6970
 Bi 0.6010    0.6010    0.3030
 Se 0.0000    0.0000    0.5000
 Se 0.2060    0.2060    0.1180
 Se 0.7940    0.7940    0.8820
end atoms_frac


begin projections 
random 
end projections


mp_grid = 4 4 4 
begin kpoints
  0.00000000   0.00000000   0.00000000 
  0.00000000   0.00000000   0.25000000 
  0.00000000   0.00000000   0.50000000 
  0.00000000   0.00000000   0.75000000 
  0.00000000   0.25000000   0.00000000 
  0.00000000   0.25000000   0.25000000 
  0.00000000   0.25000000   0.50000000 
  0.00000000   0.25000000   0.75000000 
  0.00000000   0.50000000   0.00000000 
  0.00000000   0.50000000   0.25000000 
  0.00000000   0.50000000   0.50000000 
  0.00000000   0.50000000   0.75000000 
  0.00000000   0.75000000   0.00000000 
  0.00000000   0.75000000   0.25000000 
  0.00000000   0.75000000   0.50000000 
  0.00000000   0.75000000   0.75000000 
  0.25000000   0.00000000   0.00000000 
  0.25000000   0.00000000   0.25000000 
  0.25000000   0.00000000   0.50000000 
  0.25000000   0.00000000   0.75000000 
  0.25000000   0.25000000   0.00000000 
  0.25000000   0.25000000   0.25000000 
  0.25000000   0.25000000   0.50000000 
  0.25000000   0.25000000   0.75000000 
  0.25000000   0.50000000   0.00000000 
  0.25000000   0.50000000   0.25000000 
  0.25000000   0.50000000   0.50000000 
  0.25000000   0.50000000   0.75000000 
  0.25000000   0.75000000   0.00000000 
  0.25000000   0.75000000   0.25000000 
  0.25000000   0.75000000   0.50000000 
  0.25000000   0.75000000   0.75000000 
  0.50000000   0.00000000   0.00000000 
  0.50000000   0.00000000   0.25000000 
  0.50000000   0.00000000   0.50000000 
  0.50000000   0.00000000   0.75000000 
  0.50000000   0.25000000   0.00000000 
  0.50000000   0.25000000   0.25000000 
  0.50000000   0.25000000   0.50000000 
  0.50000000   0.25000000   0.75000000 
  0.50000000   0.50000000   0.00000000 
  0.50000000   0.50000000   0.25000000 
  0.50000000   0.50000000   0.50000000 
  0.50000000   0.50000000   0.75000000 
  0.50000000   0.75000000   0.00000000 
  0.50000000   0.75000000   0.25000000 
  0.50000000   0.75000000   0.50000000 
  0.50000000   0.75000000   0.75000000 
  0.75000000   0.00000000   0.00000000 
  0.75000000   0.00000000   0.25000000 
  0.75000000   0.00000000   0.50000000 
  0.75000000   0.00000000   0.75000000 
  0.75000000   0.25000000   0.00000000 
  0.75000000   0.25000000   0.25000000 
  0.75000000   0.25000000   0.50000000 
  0.75000000   0.25000000   0.75000000 
  0.75000000   0.50000000   0.00000000 
  0.75000000   0.50000000   0.25000000 
  0.75000000   0.50000000   0.50000000 
  0.75000000   0.50000000   0.75000000 
  0.75000000   0.75000000   0.00000000 
  0.75000000   0.75000000   0.25000000 
  0.75000000   0.75000000   0.50000000 
  0.75000000   0.75000000   0.75000000 
end kpoints
\end{lstlisting}
\section{Appendix:  $\mathrm{HfP}_2$}
\subsection{HfP2.scf.in} 
\begin{lstlisting}
&CONTROL
calculation ='scf'
prefix = 'HfP2'
outdir = './bin'
pseudo_dir = './PP/'
verbosity='high'
/

&system
ibrav = 0
nat = 12
ntyp = 2
ecutwfc = 50 !Ryberg
!occupations = 'tetrahedra_opt'
occupations = 'smearing'
smearing = 'gaussian'
degauss = 0.001

!SOC
noncolin = .FALSE.
lspinorb = .FALSE.
starting_magnetization(1) = 0
starting_magnetization(2) = 0
/

&ELECTRONS
 mixing_beta = 0.495
 conv_thr =  1.0d-6
 mixing_mode = 'TF'
 diagonalization= 'david'
/

ATOMIC_SPECIES
P 30.974 P.upf
Hf 178.49 Hf.upf


ATOMIC_POSITIONS (crystal)
Hf 0.222920 0.750000 0.335788
Hf 0.777080 0.250000 0.664212
Hf 0.277080 0.250000 0.835788
Hf 0.722920 0.750000 0.164212
P 0.092791 0.750000 0.642458
P 0.907209 0.250000 0.357542
P 0.407209 0.250000 0.142458
P 0.592791 0.750000 0.857542
P 0.110700 0.750000 0.039379
P 0.889300 0.250000 0.960621
P 0.389300 0.250000 0.539379
P 0.610700 0.750000 0.460621


CELL_PARAMETERS (angstrom)
 6.460131 0.000000 0.000000
 0.000000 3.509322  0.000000
 0.000000 0.000000 8.687298


K_POINTS (automatic)
8 8 8 0 0 0

\end{lstlisting}
\subsection{HfP2.nscf.in} 
\begin{lstlisting}
&CONTROL
calculation ='nscf'
prefix = 'HfP2'
outdir = './bin'
pseudo_dir = './PP/'
verbosity='high'
/

&system
ibrav = 0
nat = 12
ntyp = 2
ecutwfc = 50 !Ryberg
occupations = 'smearing'
smearing = 'gaussian'
degauss = 0.01
nosym = .TRUE.
nbnd = 120

!SOC
noncolin = .FALSE.
lspinorb = .FALSE.
starting_magnetization(1) = 0
starting_magnetization(2) = 0
/

&ELECTRONS
 mixing_beta = 0.495
 conv_thr =  1.0d-6
 mixing_mode = 'TF'
 diagonalization= 'david'
/

ATOMIC_SPECIES
P 30.974 P.upf
Hf 178.49 Hf.upf



ATOMIC_POSITIONS (crystal)
Hf 0.222920 0.750000 0.335788
Hf 0.777080 0.250000 0.664212
Hf 0.277080 0.250000 0.835788
Hf 0.722920 0.750000 0.164212
P 0.092791 0.750000 0.642458
P 0.907209 0.250000 0.357542
P 0.407209 0.250000 0.142458
P 0.592791 0.750000 0.857542
P 0.110700 0.750000 0.039379
P 0.889300 0.250000 0.960621
P 0.389300 0.250000 0.539379
P 0.610700 0.750000 0.460621


CELL_PARAMETERS (angstrom)
 6.460131 0.000000 0.000000
 0.000000 3.509322  0.000000
 0.000000 0.000000 8.687298


K_POINTS (crystal)
512
  0.00000000   0.00000000   0.00000000 1
  0.00000000   0.00000000   0.12500000 1
  0.00000000   0.00000000   0.25000000 1
  0.00000000   0.00000000   0.37500000 1
  0.00000000   0.00000000   0.50000000 1
  0.00000000   0.00000000   0.62500000 1
  0.00000000   0.00000000   0.75000000 1
  0.00000000   0.00000000   0.87500000 1
  0.00000000   0.12500000   0.00000000 1
  0.00000000   0.12500000   0.12500000 1
  0.00000000   0.12500000   0.25000000 1
  0.00000000   0.12500000   0.37500000 1
  0.00000000   0.12500000   0.50000000 1
  0.00000000   0.12500000   0.62500000 1
  0.00000000   0.12500000   0.75000000 1
  0.00000000   0.12500000   0.87500000 1
  0.00000000   0.25000000   0.00000000 1
  0.00000000   0.25000000   0.12500000 1
  0.00000000   0.25000000   0.25000000 1
  0.00000000   0.25000000   0.37500000 1
  0.00000000   0.25000000   0.50000000 1
  0.00000000   0.25000000   0.62500000 1
  0.00000000   0.25000000   0.75000000 1
  0.00000000   0.25000000   0.87500000 1
  0.00000000   0.37500000   0.00000000 1
  0.00000000   0.37500000   0.12500000 1
  0.00000000   0.37500000   0.25000000 1
  0.00000000   0.37500000   0.37500000 1
  0.00000000   0.37500000   0.50000000 1
  0.00000000   0.37500000   0.62500000 1
  0.00000000   0.37500000   0.75000000 1
  0.00000000   0.37500000   0.87500000 1
  0.00000000   0.50000000   0.00000000 1
  0.00000000   0.50000000   0.12500000 1
  0.00000000   0.50000000   0.25000000 1
  0.00000000   0.50000000   0.37500000 1
  0.00000000   0.50000000   0.50000000 1
  0.00000000   0.50000000   0.62500000 1
  0.00000000   0.50000000   0.75000000 1
  0.00000000   0.50000000   0.87500000 1
  0.00000000   0.62500000   0.00000000 1
  0.00000000   0.62500000   0.12500000 1
  0.00000000   0.62500000   0.25000000 1
  0.00000000   0.62500000   0.37500000 1
  0.00000000   0.62500000   0.50000000 1
  0.00000000   0.62500000   0.62500000 1
  0.00000000   0.62500000   0.75000000 1
  0.00000000   0.62500000   0.87500000 1
  0.00000000   0.75000000   0.00000000 1
  0.00000000   0.75000000   0.12500000 1
  0.00000000   0.75000000   0.25000000 1
  0.00000000   0.75000000   0.37500000 1
  0.00000000   0.75000000   0.50000000 1
  0.00000000   0.75000000   0.62500000 1
  0.00000000   0.75000000   0.75000000 1
  0.00000000   0.75000000   0.87500000 1
  0.00000000   0.87500000   0.00000000 1
  0.00000000   0.87500000   0.12500000 1
  0.00000000   0.87500000   0.25000000 1
  0.00000000   0.87500000   0.37500000 1
  0.00000000   0.87500000   0.50000000 1
  0.00000000   0.87500000   0.62500000 1
  0.00000000   0.87500000   0.75000000 1
  0.00000000   0.87500000   0.87500000 1
  0.12500000   0.00000000   0.00000000 1
  0.12500000   0.00000000   0.12500000 1
  0.12500000   0.00000000   0.25000000 1
  0.12500000   0.00000000   0.37500000 1
  0.12500000   0.00000000   0.50000000 1
  0.12500000   0.00000000   0.62500000 1
  0.12500000   0.00000000   0.75000000 1
  0.12500000   0.00000000   0.87500000 1
  0.12500000   0.12500000   0.00000000 1
  0.12500000   0.12500000   0.12500000 1
  0.12500000   0.12500000   0.25000000 1
  0.12500000   0.12500000   0.37500000 1
  0.12500000   0.12500000   0.50000000 1
  0.12500000   0.12500000   0.62500000 1
  0.12500000   0.12500000   0.75000000 1
  0.12500000   0.12500000   0.87500000 1
  0.12500000   0.25000000   0.00000000 1
  0.12500000   0.25000000   0.12500000 1
  0.12500000   0.25000000   0.25000000 1
  0.12500000   0.25000000   0.37500000 1
  0.12500000   0.25000000   0.50000000 1
  0.12500000   0.25000000   0.62500000 1
  0.12500000   0.25000000   0.75000000 1
  0.12500000   0.25000000   0.87500000 1
  0.12500000   0.37500000   0.00000000 1
  0.12500000   0.37500000   0.12500000 1
  0.12500000   0.37500000   0.25000000 1
  0.12500000   0.37500000   0.37500000 1
  0.12500000   0.37500000   0.50000000 1
  0.12500000   0.37500000   0.62500000 1
  0.12500000   0.37500000   0.75000000 1
  0.12500000   0.37500000   0.87500000 1
  0.12500000   0.50000000   0.00000000 1
  0.12500000   0.50000000   0.12500000 1
  0.12500000   0.50000000   0.25000000 1
  0.12500000   0.50000000   0.37500000 1
  0.12500000   0.50000000   0.50000000 1
  0.12500000   0.50000000   0.62500000 1
  0.12500000   0.50000000   0.75000000 1
  0.12500000   0.50000000   0.87500000 1
  0.12500000   0.62500000   0.00000000 1
  0.12500000   0.62500000   0.12500000 1
  0.12500000   0.62500000   0.25000000 1
  0.12500000   0.62500000   0.37500000 1
  0.12500000   0.62500000   0.50000000 1
  0.12500000   0.62500000   0.62500000 1
  0.12500000   0.62500000   0.75000000 1
  0.12500000   0.62500000   0.87500000 1
  0.12500000   0.75000000   0.00000000 1
  0.12500000   0.75000000   0.12500000 1
  0.12500000   0.75000000   0.25000000 1
  0.12500000   0.75000000   0.37500000 1
  0.12500000   0.75000000   0.50000000 1
  0.12500000   0.75000000   0.62500000 1
  0.12500000   0.75000000   0.75000000 1
  0.12500000   0.75000000   0.87500000 1
  0.12500000   0.87500000   0.00000000 1
  0.12500000   0.87500000   0.12500000 1
  0.12500000   0.87500000   0.25000000 1
  0.12500000   0.87500000   0.37500000 1
  0.12500000   0.87500000   0.50000000 1
  0.12500000   0.87500000   0.62500000 1
  0.12500000   0.87500000   0.75000000 1
  0.12500000   0.87500000   0.87500000 1
  0.25000000   0.00000000   0.00000000 1
  0.25000000   0.00000000   0.12500000 1
  0.25000000   0.00000000   0.25000000 1
  0.25000000   0.00000000   0.37500000 1
  0.25000000   0.00000000   0.50000000 1
  0.25000000   0.00000000   0.62500000 1
  0.25000000   0.00000000   0.75000000 1
  0.25000000   0.00000000   0.87500000 1
  0.25000000   0.12500000   0.00000000 1
  0.25000000   0.12500000   0.12500000 1
  0.25000000   0.12500000   0.25000000 1
  0.25000000   0.12500000   0.37500000 1
  0.25000000   0.12500000   0.50000000 1
  0.25000000   0.12500000   0.62500000 1
  0.25000000   0.12500000   0.75000000 1
  0.25000000   0.12500000   0.87500000 1
  0.25000000   0.25000000   0.00000000 1
  0.25000000   0.25000000   0.12500000 1
  0.25000000   0.25000000   0.25000000 1
  0.25000000   0.25000000   0.37500000 1
  0.25000000   0.25000000   0.50000000 1
  0.25000000   0.25000000   0.62500000 1
  0.25000000   0.25000000   0.75000000 1
  0.25000000   0.25000000   0.87500000 1
  0.25000000   0.37500000   0.00000000 1
  0.25000000   0.37500000   0.12500000 1
  0.25000000   0.37500000   0.25000000 1
  0.25000000   0.37500000   0.37500000 1
  0.25000000   0.37500000   0.50000000 1
  0.25000000   0.37500000   0.62500000 1
  0.25000000   0.37500000   0.75000000 1
  0.25000000   0.37500000   0.87500000 1
  0.25000000   0.50000000   0.00000000 1
  0.25000000   0.50000000   0.12500000 1
  0.25000000   0.50000000   0.25000000 1
  0.25000000   0.50000000   0.37500000 1
  0.25000000   0.50000000   0.50000000 1
  0.25000000   0.50000000   0.62500000 1
  0.25000000   0.50000000   0.75000000 1
  0.25000000   0.50000000   0.87500000 1
  0.25000000   0.62500000   0.00000000 1
  0.25000000   0.62500000   0.12500000 1
  0.25000000   0.62500000   0.25000000 1
  0.25000000   0.62500000   0.37500000 1
  0.25000000   0.62500000   0.50000000 1
  0.25000000   0.62500000   0.62500000 1
  0.25000000   0.62500000   0.75000000 1
  0.25000000   0.62500000   0.87500000 1
  0.25000000   0.75000000   0.00000000 1
  0.25000000   0.75000000   0.12500000 1
  0.25000000   0.75000000   0.25000000 1
  0.25000000   0.75000000   0.37500000 1
  0.25000000   0.75000000   0.50000000 1
  0.25000000   0.75000000   0.62500000 1
  0.25000000   0.75000000   0.75000000 1
  0.25000000   0.75000000   0.87500000 1
  0.25000000   0.87500000   0.00000000 1
  0.25000000   0.87500000   0.12500000 1
  0.25000000   0.87500000   0.25000000 1
  0.25000000   0.87500000   0.37500000 1
  0.25000000   0.87500000   0.50000000 1
  0.25000000   0.87500000   0.62500000 1
  0.25000000   0.87500000   0.75000000 1
  0.25000000   0.87500000   0.87500000 1
  0.37500000   0.00000000   0.00000000 1
  0.37500000   0.00000000   0.12500000 1
  0.37500000   0.00000000   0.25000000 1
  0.37500000   0.00000000   0.37500000 1
  0.37500000   0.00000000   0.50000000 1
  0.37500000   0.00000000   0.62500000 1
  0.37500000   0.00000000   0.75000000 1
  0.37500000   0.00000000   0.87500000 1
  0.37500000   0.12500000   0.00000000 1
  0.37500000   0.12500000   0.12500000 1
  0.37500000   0.12500000   0.25000000 1
  0.37500000   0.12500000   0.37500000 1
  0.37500000   0.12500000   0.50000000 1
  0.37500000   0.12500000   0.62500000 1
  0.37500000   0.12500000   0.75000000 1
  0.37500000   0.12500000   0.87500000 1
  0.37500000   0.25000000   0.00000000 1
  0.37500000   0.25000000   0.12500000 1
  0.37500000   0.25000000   0.25000000 1
  0.37500000   0.25000000   0.37500000 1
  0.37500000   0.25000000   0.50000000 1
  0.37500000   0.25000000   0.62500000 1
  0.37500000   0.25000000   0.75000000 1
  0.37500000   0.25000000   0.87500000 1
  0.37500000   0.37500000   0.00000000 1
  0.37500000   0.37500000   0.12500000 1
  0.37500000   0.37500000   0.25000000 1
  0.37500000   0.37500000   0.37500000 1
  0.37500000   0.37500000   0.50000000 1
  0.37500000   0.37500000   0.62500000 1
  0.37500000   0.37500000   0.75000000 1
  0.37500000   0.37500000   0.87500000 1
  0.37500000   0.50000000   0.00000000 1
  0.37500000   0.50000000   0.12500000 1
  0.37500000   0.50000000   0.25000000 1
  0.37500000   0.50000000   0.37500000 1
  0.37500000   0.50000000   0.50000000 1
  0.37500000   0.50000000   0.62500000 1
  0.37500000   0.50000000   0.75000000 1
  0.37500000   0.50000000   0.87500000 1
  0.37500000   0.62500000   0.00000000 1
  0.37500000   0.62500000   0.12500000 1
  0.37500000   0.62500000   0.25000000 1
  0.37500000   0.62500000   0.37500000 1
  0.37500000   0.62500000   0.50000000 1
  0.37500000   0.62500000   0.62500000 1
  0.37500000   0.62500000   0.75000000 1
  0.37500000   0.62500000   0.87500000 1
  0.37500000   0.75000000   0.00000000 1
  0.37500000   0.75000000   0.12500000 1
  0.37500000   0.75000000   0.25000000 1
  0.37500000   0.75000000   0.37500000 1
  0.37500000   0.75000000   0.50000000 1
  0.37500000   0.75000000   0.62500000 1
  0.37500000   0.75000000   0.75000000 1
  0.37500000   0.75000000   0.87500000 1
  0.37500000   0.87500000   0.00000000 1
  0.37500000   0.87500000   0.12500000 1
  0.37500000   0.87500000   0.25000000 1
  0.37500000   0.87500000   0.37500000 1
  0.37500000   0.87500000   0.50000000 1
  0.37500000   0.87500000   0.62500000 1
  0.37500000   0.87500000   0.75000000 1
  0.37500000   0.87500000   0.87500000 1
  0.50000000   0.00000000   0.00000000 1
  0.50000000   0.00000000   0.12500000 1
  0.50000000   0.00000000   0.25000000 1
  0.50000000   0.00000000   0.37500000 1
  0.50000000   0.00000000   0.50000000 1
  0.50000000   0.00000000   0.62500000 1
  0.50000000   0.00000000   0.75000000 1
  0.50000000   0.00000000   0.87500000 1
  0.50000000   0.12500000   0.00000000 1
  0.50000000   0.12500000   0.12500000 1
  0.50000000   0.12500000   0.25000000 1
  0.50000000   0.12500000   0.37500000 1
  0.50000000   0.12500000   0.50000000 1
  0.50000000   0.12500000   0.62500000 1
  0.50000000   0.12500000   0.75000000 1
  0.50000000   0.12500000   0.87500000 1
  0.50000000   0.25000000   0.00000000 1
  0.50000000   0.25000000   0.12500000 1
  0.50000000   0.25000000   0.25000000 1
  0.50000000   0.25000000   0.37500000 1
  0.50000000   0.25000000   0.50000000 1
  0.50000000   0.25000000   0.62500000 1
  0.50000000   0.25000000   0.75000000 1
  0.50000000   0.25000000   0.87500000 1
  0.50000000   0.37500000   0.00000000 1
  0.50000000   0.37500000   0.12500000 1
  0.50000000   0.37500000   0.25000000 1
  0.50000000   0.37500000   0.37500000 1
  0.50000000   0.37500000   0.50000000 1
  0.50000000   0.37500000   0.62500000 1
  0.50000000   0.37500000   0.75000000 1
  0.50000000   0.37500000   0.87500000 1
  0.50000000   0.50000000   0.00000000 1
  0.50000000   0.50000000   0.12500000 1
  0.50000000   0.50000000   0.25000000 1
  0.50000000   0.50000000   0.37500000 1
  0.50000000   0.50000000   0.50000000 1
  0.50000000   0.50000000   0.62500000 1
  0.50000000   0.50000000   0.75000000 1
  0.50000000   0.50000000   0.87500000 1
  0.50000000   0.62500000   0.00000000 1
  0.50000000   0.62500000   0.12500000 1
  0.50000000   0.62500000   0.25000000 1
  0.50000000   0.62500000   0.37500000 1
  0.50000000   0.62500000   0.50000000 1
  0.50000000   0.62500000   0.62500000 1
  0.50000000   0.62500000   0.75000000 1
  0.50000000   0.62500000   0.87500000 1
  0.50000000   0.75000000   0.00000000 1
  0.50000000   0.75000000   0.12500000 1
  0.50000000   0.75000000   0.25000000 1
  0.50000000   0.75000000   0.37500000 1
  0.50000000   0.75000000   0.50000000 1
  0.50000000   0.75000000   0.62500000 1
  0.50000000   0.75000000   0.75000000 1
  0.50000000   0.75000000   0.87500000 1
  0.50000000   0.87500000   0.00000000 1
  0.50000000   0.87500000   0.12500000 1
  0.50000000   0.87500000   0.25000000 1
  0.50000000   0.87500000   0.37500000 1
  0.50000000   0.87500000   0.50000000 1
  0.50000000   0.87500000   0.62500000 1
  0.50000000   0.87500000   0.75000000 1
  0.50000000   0.87500000   0.87500000 1
  0.62500000   0.00000000   0.00000000 1
  0.62500000   0.00000000   0.12500000 1
  0.62500000   0.00000000   0.25000000 1
  0.62500000   0.00000000   0.37500000 1
  0.62500000   0.00000000   0.50000000 1
  0.62500000   0.00000000   0.62500000 1
  0.62500000   0.00000000   0.75000000 1
  0.62500000   0.00000000   0.87500000 1
  0.62500000   0.12500000   0.00000000 1
  0.62500000   0.12500000   0.12500000 1
  0.62500000   0.12500000   0.25000000 1
  0.62500000   0.12500000   0.37500000 1
  0.62500000   0.12500000   0.50000000 1
  0.62500000   0.12500000   0.62500000 1
  0.62500000   0.12500000   0.75000000 1
  0.62500000   0.12500000   0.87500000 1
  0.62500000   0.25000000   0.00000000 1
  0.62500000   0.25000000   0.12500000 1
  0.62500000   0.25000000   0.25000000 1
  0.62500000   0.25000000   0.37500000 1
  0.62500000   0.25000000   0.50000000 1
  0.62500000   0.25000000   0.62500000 1
  0.62500000   0.25000000   0.75000000 1
  0.62500000   0.25000000   0.87500000 1
  0.62500000   0.37500000   0.00000000 1
  0.62500000   0.37500000   0.12500000 1
  0.62500000   0.37500000   0.25000000 1
  0.62500000   0.37500000   0.37500000 1
  0.62500000   0.37500000   0.50000000 1
  0.62500000   0.37500000   0.62500000 1
  0.62500000   0.37500000   0.75000000 1
  0.62500000   0.37500000   0.87500000 1
  0.62500000   0.50000000   0.00000000 1
  0.62500000   0.50000000   0.12500000 1
  0.62500000   0.50000000   0.25000000 1
  0.62500000   0.50000000   0.37500000 1
  0.62500000   0.50000000   0.50000000 1
  0.62500000   0.50000000   0.62500000 1
  0.62500000   0.50000000   0.75000000 1
  0.62500000   0.50000000   0.87500000 1
  0.62500000   0.62500000   0.00000000 1
  0.62500000   0.62500000   0.12500000 1
  0.62500000   0.62500000   0.25000000 1
  0.62500000   0.62500000   0.37500000 1
  0.62500000   0.62500000   0.50000000 1
  0.62500000   0.62500000   0.62500000 1
  0.62500000   0.62500000   0.75000000 1
  0.62500000   0.62500000   0.87500000 1
  0.62500000   0.75000000   0.00000000 1
  0.62500000   0.75000000   0.12500000 1
  0.62500000   0.75000000   0.25000000 1
  0.62500000   0.75000000   0.37500000 1
  0.62500000   0.75000000   0.50000000 1
  0.62500000   0.75000000   0.62500000 1
  0.62500000   0.75000000   0.75000000 1
  0.62500000   0.75000000   0.87500000 1
  0.62500000   0.87500000   0.00000000 1
  0.62500000   0.87500000   0.12500000 1
  0.62500000   0.87500000   0.25000000 1
  0.62500000   0.87500000   0.37500000 1
  0.62500000   0.87500000   0.50000000 1
  0.62500000   0.87500000   0.62500000 1
  0.62500000   0.87500000   0.75000000 1
  0.62500000   0.87500000   0.87500000 1
  0.75000000   0.00000000   0.00000000 1
  0.75000000   0.00000000   0.12500000 1
  0.75000000   0.00000000   0.25000000 1
  0.75000000   0.00000000   0.37500000 1
  0.75000000   0.00000000   0.50000000 1
  0.75000000   0.00000000   0.62500000 1
  0.75000000   0.00000000   0.75000000 1
  0.75000000   0.00000000   0.87500000 1
  0.75000000   0.12500000   0.00000000 1
  0.75000000   0.12500000   0.12500000 1
  0.75000000   0.12500000   0.25000000 1
  0.75000000   0.12500000   0.37500000 1
  0.75000000   0.12500000   0.50000000 1
  0.75000000   0.12500000   0.62500000 1
  0.75000000   0.12500000   0.75000000 1
  0.75000000   0.12500000   0.87500000 1
  0.75000000   0.25000000   0.00000000 1
  0.75000000   0.25000000   0.12500000 1
  0.75000000   0.25000000   0.25000000 1
  0.75000000   0.25000000   0.37500000 1
  0.75000000   0.25000000   0.50000000 1
  0.75000000   0.25000000   0.62500000 1
  0.75000000   0.25000000   0.75000000 1
  0.75000000   0.25000000   0.87500000 1
  0.75000000   0.37500000   0.00000000 1
  0.75000000   0.37500000   0.12500000 1
  0.75000000   0.37500000   0.25000000 1
  0.75000000   0.37500000   0.37500000 1
  0.75000000   0.37500000   0.50000000 1
  0.75000000   0.37500000   0.62500000 1
  0.75000000   0.37500000   0.75000000 1
  0.75000000   0.37500000   0.87500000 1
  0.75000000   0.50000000   0.00000000 1
  0.75000000   0.50000000   0.12500000 1
  0.75000000   0.50000000   0.25000000 1
  0.75000000   0.50000000   0.37500000 1
  0.75000000   0.50000000   0.50000000 1
  0.75000000   0.50000000   0.62500000 1
  0.75000000   0.50000000   0.75000000 1
  0.75000000   0.50000000   0.87500000 1
  0.75000000   0.62500000   0.00000000 1
  0.75000000   0.62500000   0.12500000 1
  0.75000000   0.62500000   0.25000000 1
  0.75000000   0.62500000   0.37500000 1
  0.75000000   0.62500000   0.50000000 1
  0.75000000   0.62500000   0.62500000 1
  0.75000000   0.62500000   0.75000000 1
  0.75000000   0.62500000   0.87500000 1
  0.75000000   0.75000000   0.00000000 1
  0.75000000   0.75000000   0.12500000 1
  0.75000000   0.75000000   0.25000000 1
  0.75000000   0.75000000   0.37500000 1
  0.75000000   0.75000000   0.50000000 1
  0.75000000   0.75000000   0.62500000 1
  0.75000000   0.75000000   0.75000000 1
  0.75000000   0.75000000   0.87500000 1
  0.75000000   0.87500000   0.00000000 1
  0.75000000   0.87500000   0.12500000 1
  0.75000000   0.87500000   0.25000000 1
  0.75000000   0.87500000   0.37500000 1
  0.75000000   0.87500000   0.50000000 1
  0.75000000   0.87500000   0.62500000 1
  0.75000000   0.87500000   0.75000000 1
  0.75000000   0.87500000   0.87500000 1
  0.87500000   0.00000000   0.00000000 1
  0.87500000   0.00000000   0.12500000 1
  0.87500000   0.00000000   0.25000000 1
  0.87500000   0.00000000   0.37500000 1
  0.87500000   0.00000000   0.50000000 1
  0.87500000   0.00000000   0.62500000 1
  0.87500000   0.00000000   0.75000000 1
  0.87500000   0.00000000   0.87500000 1
  0.87500000   0.12500000   0.00000000 1
  0.87500000   0.12500000   0.12500000 1
  0.87500000   0.12500000   0.25000000 1
  0.87500000   0.12500000   0.37500000 1
  0.87500000   0.12500000   0.50000000 1
  0.87500000   0.12500000   0.62500000 1
  0.87500000   0.12500000   0.75000000 1
  0.87500000   0.12500000   0.87500000 1
  0.87500000   0.25000000   0.00000000 1
  0.87500000   0.25000000   0.12500000 1
  0.87500000   0.25000000   0.25000000 1
  0.87500000   0.25000000   0.37500000 1
  0.87500000   0.25000000   0.50000000 1
  0.87500000   0.25000000   0.62500000 1
  0.87500000   0.25000000   0.75000000 1
  0.87500000   0.25000000   0.87500000 1
  0.87500000   0.37500000   0.00000000 1
  0.87500000   0.37500000   0.12500000 1
  0.87500000   0.37500000   0.25000000 1
  0.87500000   0.37500000   0.37500000 1
  0.87500000   0.37500000   0.50000000 1
  0.87500000   0.37500000   0.62500000 1
  0.87500000   0.37500000   0.75000000 1
  0.87500000   0.37500000   0.87500000 1
  0.87500000   0.50000000   0.00000000 1
  0.87500000   0.50000000   0.12500000 1
  0.87500000   0.50000000   0.25000000 1
  0.87500000   0.50000000   0.37500000 1
  0.87500000   0.50000000   0.50000000 1
  0.87500000   0.50000000   0.62500000 1
  0.87500000   0.50000000   0.75000000 1
  0.87500000   0.50000000   0.87500000 1
  0.87500000   0.62500000   0.00000000 1
  0.87500000   0.62500000   0.12500000 1
  0.87500000   0.62500000   0.25000000 1
  0.87500000   0.62500000   0.37500000 1
  0.87500000   0.62500000   0.50000000 1
  0.87500000   0.62500000   0.62500000 1
  0.87500000   0.62500000   0.75000000 1
  0.87500000   0.62500000   0.87500000 1
  0.87500000   0.75000000   0.00000000 1
  0.87500000   0.75000000   0.12500000 1
  0.87500000   0.75000000   0.25000000 1
  0.87500000   0.75000000   0.37500000 1
  0.87500000   0.75000000   0.50000000 1
  0.87500000   0.75000000   0.62500000 1
  0.87500000   0.75000000   0.75000000 1
  0.87500000   0.75000000   0.87500000 1
  0.87500000   0.87500000   0.00000000 1
  0.87500000   0.87500000   0.12500000 1
  0.87500000   0.87500000   0.25000000 1
  0.87500000   0.87500000   0.37500000 1
  0.87500000   0.87500000   0.50000000 1
  0.87500000   0.87500000   0.62500000 1
  0.87500000   0.87500000   0.75000000 1
  0.87500000   0.87500000   0.87500000 1

\end{lstlisting}
\subsection{HfP2.win} 
\begin{lstlisting}
write_hr = true
write_xyz = true
spinors = false
num_wann = 120
num_iter = 500
!trial_step = 100

!!! 13.3711
!min of outer window
!dis_win_min = 10.0
!dis_win_max = 16.0

!inner
!dis_froz_min = 12
!dis_froz_max = 14

begin unit_cell_cart
 6.460131 0.000000 0.000000
 0.000000 3.509322  0.000000
 0.000000 0.000000 8.687298
end unit_cell_cart


begin atoms_frac
Hf 0.222920 0.750000 0.335788
Hf 0.777080 0.250000 0.664212
Hf 0.277080 0.250000 0.835788
Hf 0.722920 0.750000 0.164212
P 0.092791 0.750000 0.642458
P 0.907209 0.250000 0.357542
P 0.407209 0.250000 0.142458
P 0.592791 0.750000 0.857542
P 0.110700 0.750000 0.039379
P 0.889300 0.250000 0.960621
P 0.389300 0.250000 0.539379
P 0.610700 0.750000 0.460621
end atoms_frac


begin projections 
random 
!Hf:s;dz2;dxz;dyz;dxy
!P:s;pz;px;py
end projections

mp_grid = 8 8 8 
begin kpoints
  0.00000000   0.00000000   0.00000000 
  0.00000000   0.00000000   0.12500000 
  0.00000000   0.00000000   0.25000000 
  0.00000000   0.00000000   0.37500000 
  0.00000000   0.00000000   0.50000000 
  0.00000000   0.00000000   0.62500000 
  0.00000000   0.00000000   0.75000000 
  0.00000000   0.00000000   0.87500000 
  0.00000000   0.12500000   0.00000000 
  0.00000000   0.12500000   0.12500000 
  0.00000000   0.12500000   0.25000000 
  0.00000000   0.12500000   0.37500000 
  0.00000000   0.12500000   0.50000000 
  0.00000000   0.12500000   0.62500000 
  0.00000000   0.12500000   0.75000000 
  0.00000000   0.12500000   0.87500000 
  0.00000000   0.25000000   0.00000000 
  0.00000000   0.25000000   0.12500000 
  0.00000000   0.25000000   0.25000000 
  0.00000000   0.25000000   0.37500000 
  0.00000000   0.25000000   0.50000000 
  0.00000000   0.25000000   0.62500000 
  0.00000000   0.25000000   0.75000000 
  0.00000000   0.25000000   0.87500000 
  0.00000000   0.37500000   0.00000000 
  0.00000000   0.37500000   0.12500000 
  0.00000000   0.37500000   0.25000000 
  0.00000000   0.37500000   0.37500000 
  0.00000000   0.37500000   0.50000000 
  0.00000000   0.37500000   0.62500000 
  0.00000000   0.37500000   0.75000000 
  0.00000000   0.37500000   0.87500000 
  0.00000000   0.50000000   0.00000000 
  0.00000000   0.50000000   0.12500000 
  0.00000000   0.50000000   0.25000000 
  0.00000000   0.50000000   0.37500000 
  0.00000000   0.50000000   0.50000000 
  0.00000000   0.50000000   0.62500000 
  0.00000000   0.50000000   0.75000000 
  0.00000000   0.50000000   0.87500000 
  0.00000000   0.62500000   0.00000000 
  0.00000000   0.62500000   0.12500000 
  0.00000000   0.62500000   0.25000000 
  0.00000000   0.62500000   0.37500000 
  0.00000000   0.62500000   0.50000000 
  0.00000000   0.62500000   0.62500000 
  0.00000000   0.62500000   0.75000000 
  0.00000000   0.62500000   0.87500000 
  0.00000000   0.75000000   0.00000000 
  0.00000000   0.75000000   0.12500000 
  0.00000000   0.75000000   0.25000000 
  0.00000000   0.75000000   0.37500000 
  0.00000000   0.75000000   0.50000000 
  0.00000000   0.75000000   0.62500000 
  0.00000000   0.75000000   0.75000000 
  0.00000000   0.75000000   0.87500000 
  0.00000000   0.87500000   0.00000000 
  0.00000000   0.87500000   0.12500000 
  0.00000000   0.87500000   0.25000000 
  0.00000000   0.87500000   0.37500000 
  0.00000000   0.87500000   0.50000000 
  0.00000000   0.87500000   0.62500000 
  0.00000000   0.87500000   0.75000000 
  0.00000000   0.87500000   0.87500000 
  0.12500000   0.00000000   0.00000000 
  0.12500000   0.00000000   0.12500000 
  0.12500000   0.00000000   0.25000000 
  0.12500000   0.00000000   0.37500000 
  0.12500000   0.00000000   0.50000000 
  0.12500000   0.00000000   0.62500000 
  0.12500000   0.00000000   0.75000000 
  0.12500000   0.00000000   0.87500000 
  0.12500000   0.12500000   0.00000000 
  0.12500000   0.12500000   0.12500000 
  0.12500000   0.12500000   0.25000000 
  0.12500000   0.12500000   0.37500000 
  0.12500000   0.12500000   0.50000000 
  0.12500000   0.12500000   0.62500000 
  0.12500000   0.12500000   0.75000000 
  0.12500000   0.12500000   0.87500000 
  0.12500000   0.25000000   0.00000000 
  0.12500000   0.25000000   0.12500000 
  0.12500000   0.25000000   0.25000000 
  0.12500000   0.25000000   0.37500000 
  0.12500000   0.25000000   0.50000000 
  0.12500000   0.25000000   0.62500000 
  0.12500000   0.25000000   0.75000000 
  0.12500000   0.25000000   0.87500000 
  0.12500000   0.37500000   0.00000000 
  0.12500000   0.37500000   0.12500000 
  0.12500000   0.37500000   0.25000000 
  0.12500000   0.37500000   0.37500000 
  0.12500000   0.37500000   0.50000000 
  0.12500000   0.37500000   0.62500000 
  0.12500000   0.37500000   0.75000000 
  0.12500000   0.37500000   0.87500000 
  0.12500000   0.50000000   0.00000000 
  0.12500000   0.50000000   0.12500000 
  0.12500000   0.50000000   0.25000000 
  0.12500000   0.50000000   0.37500000 
  0.12500000   0.50000000   0.50000000 
  0.12500000   0.50000000   0.62500000 
  0.12500000   0.50000000   0.75000000 
  0.12500000   0.50000000   0.87500000 
  0.12500000   0.62500000   0.00000000 
  0.12500000   0.62500000   0.12500000 
  0.12500000   0.62500000   0.25000000 
  0.12500000   0.62500000   0.37500000 
  0.12500000   0.62500000   0.50000000 
  0.12500000   0.62500000   0.62500000 
  0.12500000   0.62500000   0.75000000 
  0.12500000   0.62500000   0.87500000 
  0.12500000   0.75000000   0.00000000 
  0.12500000   0.75000000   0.12500000 
  0.12500000   0.75000000   0.25000000 
  0.12500000   0.75000000   0.37500000 
  0.12500000   0.75000000   0.50000000 
  0.12500000   0.75000000   0.62500000 
  0.12500000   0.75000000   0.75000000 
  0.12500000   0.75000000   0.87500000 
  0.12500000   0.87500000   0.00000000 
  0.12500000   0.87500000   0.12500000 
  0.12500000   0.87500000   0.25000000 
  0.12500000   0.87500000   0.37500000 
  0.12500000   0.87500000   0.50000000 
  0.12500000   0.87500000   0.62500000 
  0.12500000   0.87500000   0.75000000 
  0.12500000   0.87500000   0.87500000 
  0.25000000   0.00000000   0.00000000 
  0.25000000   0.00000000   0.12500000 
  0.25000000   0.00000000   0.25000000 
  0.25000000   0.00000000   0.37500000 
  0.25000000   0.00000000   0.50000000 
  0.25000000   0.00000000   0.62500000 
  0.25000000   0.00000000   0.75000000 
  0.25000000   0.00000000   0.87500000 
  0.25000000   0.12500000   0.00000000 
  0.25000000   0.12500000   0.12500000 
  0.25000000   0.12500000   0.25000000 
  0.25000000   0.12500000   0.37500000 
  0.25000000   0.12500000   0.50000000 
  0.25000000   0.12500000   0.62500000 
  0.25000000   0.12500000   0.75000000 
  0.25000000   0.12500000   0.87500000 
  0.25000000   0.25000000   0.00000000 
  0.25000000   0.25000000   0.12500000 
  0.25000000   0.25000000   0.25000000 
  0.25000000   0.25000000   0.37500000 
  0.25000000   0.25000000   0.50000000 
  0.25000000   0.25000000   0.62500000 
  0.25000000   0.25000000   0.75000000 
  0.25000000   0.25000000   0.87500000 
  0.25000000   0.37500000   0.00000000 
  0.25000000   0.37500000   0.12500000 
  0.25000000   0.37500000   0.25000000 
  0.25000000   0.37500000   0.37500000 
  0.25000000   0.37500000   0.50000000 
  0.25000000   0.37500000   0.62500000 
  0.25000000   0.37500000   0.75000000 
  0.25000000   0.37500000   0.87500000 
  0.25000000   0.50000000   0.00000000 
  0.25000000   0.50000000   0.12500000 
  0.25000000   0.50000000   0.25000000 
  0.25000000   0.50000000   0.37500000 
  0.25000000   0.50000000   0.50000000 
  0.25000000   0.50000000   0.62500000 
  0.25000000   0.50000000   0.75000000 
  0.25000000   0.50000000   0.87500000 
  0.25000000   0.62500000   0.00000000 
  0.25000000   0.62500000   0.12500000 
  0.25000000   0.62500000   0.25000000 
  0.25000000   0.62500000   0.37500000 
  0.25000000   0.62500000   0.50000000 
  0.25000000   0.62500000   0.62500000 
  0.25000000   0.62500000   0.75000000 
  0.25000000   0.62500000   0.87500000 
  0.25000000   0.75000000   0.00000000 
  0.25000000   0.75000000   0.12500000 
  0.25000000   0.75000000   0.25000000 
  0.25000000   0.75000000   0.37500000 
  0.25000000   0.75000000   0.50000000 
  0.25000000   0.75000000   0.62500000 
  0.25000000   0.75000000   0.75000000 
  0.25000000   0.75000000   0.87500000 
  0.25000000   0.87500000   0.00000000 
  0.25000000   0.87500000   0.12500000 
  0.25000000   0.87500000   0.25000000 
  0.25000000   0.87500000   0.37500000 
  0.25000000   0.87500000   0.50000000 
  0.25000000   0.87500000   0.62500000 
  0.25000000   0.87500000   0.75000000 
  0.25000000   0.87500000   0.87500000 
  0.37500000   0.00000000   0.00000000 
  0.37500000   0.00000000   0.12500000 
  0.37500000   0.00000000   0.25000000 
  0.37500000   0.00000000   0.37500000 
  0.37500000   0.00000000   0.50000000 
  0.37500000   0.00000000   0.62500000 
  0.37500000   0.00000000   0.75000000 
  0.37500000   0.00000000   0.87500000 
  0.37500000   0.12500000   0.00000000 
  0.37500000   0.12500000   0.12500000 
  0.37500000   0.12500000   0.25000000 
  0.37500000   0.12500000   0.37500000 
  0.37500000   0.12500000   0.50000000 
  0.37500000   0.12500000   0.62500000 
  0.37500000   0.12500000   0.75000000 
  0.37500000   0.12500000   0.87500000 
  0.37500000   0.25000000   0.00000000 
  0.37500000   0.25000000   0.12500000 
  0.37500000   0.25000000   0.25000000 
  0.37500000   0.25000000   0.37500000 
  0.37500000   0.25000000   0.50000000 
  0.37500000   0.25000000   0.62500000 
  0.37500000   0.25000000   0.75000000 
  0.37500000   0.25000000   0.87500000 
  0.37500000   0.37500000   0.00000000 
  0.37500000   0.37500000   0.12500000 
  0.37500000   0.37500000   0.25000000 
  0.37500000   0.37500000   0.37500000 
  0.37500000   0.37500000   0.50000000 
  0.37500000   0.37500000   0.62500000 
  0.37500000   0.37500000   0.75000000 
  0.37500000   0.37500000   0.87500000 
  0.37500000   0.50000000   0.00000000 
  0.37500000   0.50000000   0.12500000 
  0.37500000   0.50000000   0.25000000 
  0.37500000   0.50000000   0.37500000 
  0.37500000   0.50000000   0.50000000 
  0.37500000   0.50000000   0.62500000 
  0.37500000   0.50000000   0.75000000 
  0.37500000   0.50000000   0.87500000 
  0.37500000   0.62500000   0.00000000 
  0.37500000   0.62500000   0.12500000 
  0.37500000   0.62500000   0.25000000 
  0.37500000   0.62500000   0.37500000 
  0.37500000   0.62500000   0.50000000 
  0.37500000   0.62500000   0.62500000 
  0.37500000   0.62500000   0.75000000 
  0.37500000   0.62500000   0.87500000 
  0.37500000   0.75000000   0.00000000 
  0.37500000   0.75000000   0.12500000 
  0.37500000   0.75000000   0.25000000 
  0.37500000   0.75000000   0.37500000 
  0.37500000   0.75000000   0.50000000 
  0.37500000   0.75000000   0.62500000 
  0.37500000   0.75000000   0.75000000 
  0.37500000   0.75000000   0.87500000 
  0.37500000   0.87500000   0.00000000 
  0.37500000   0.87500000   0.12500000 
  0.37500000   0.87500000   0.25000000 
  0.37500000   0.87500000   0.37500000 
  0.37500000   0.87500000   0.50000000 
  0.37500000   0.87500000   0.62500000 
  0.37500000   0.87500000   0.75000000 
  0.37500000   0.87500000   0.87500000 
  0.50000000   0.00000000   0.00000000 
  0.50000000   0.00000000   0.12500000 
  0.50000000   0.00000000   0.25000000 
  0.50000000   0.00000000   0.37500000 
  0.50000000   0.00000000   0.50000000 
  0.50000000   0.00000000   0.62500000 
  0.50000000   0.00000000   0.75000000 
  0.50000000   0.00000000   0.87500000 
  0.50000000   0.12500000   0.00000000 
  0.50000000   0.12500000   0.12500000 
  0.50000000   0.12500000   0.25000000 
  0.50000000   0.12500000   0.37500000 
  0.50000000   0.12500000   0.50000000 
  0.50000000   0.12500000   0.62500000 
  0.50000000   0.12500000   0.75000000 
  0.50000000   0.12500000   0.87500000 
  0.50000000   0.25000000   0.00000000 
  0.50000000   0.25000000   0.12500000 
  0.50000000   0.25000000   0.25000000 
  0.50000000   0.25000000   0.37500000 
  0.50000000   0.25000000   0.50000000 
  0.50000000   0.25000000   0.62500000 
  0.50000000   0.25000000   0.75000000 
  0.50000000   0.25000000   0.87500000 
  0.50000000   0.37500000   0.00000000 
  0.50000000   0.37500000   0.12500000 
  0.50000000   0.37500000   0.25000000 
  0.50000000   0.37500000   0.37500000 
  0.50000000   0.37500000   0.50000000 
  0.50000000   0.37500000   0.62500000 
  0.50000000   0.37500000   0.75000000 
  0.50000000   0.37500000   0.87500000 
  0.50000000   0.50000000   0.00000000 
  0.50000000   0.50000000   0.12500000 
  0.50000000   0.50000000   0.25000000 
  0.50000000   0.50000000   0.37500000 
  0.50000000   0.50000000   0.50000000 
  0.50000000   0.50000000   0.62500000 
  0.50000000   0.50000000   0.75000000 
  0.50000000   0.50000000   0.87500000 
  0.50000000   0.62500000   0.00000000 
  0.50000000   0.62500000   0.12500000 
  0.50000000   0.62500000   0.25000000 
  0.50000000   0.62500000   0.37500000 
  0.50000000   0.62500000   0.50000000 
  0.50000000   0.62500000   0.62500000 
  0.50000000   0.62500000   0.75000000 
  0.50000000   0.62500000   0.87500000 
  0.50000000   0.75000000   0.00000000 
  0.50000000   0.75000000   0.12500000 
  0.50000000   0.75000000   0.25000000 
  0.50000000   0.75000000   0.37500000 
  0.50000000   0.75000000   0.50000000 
  0.50000000   0.75000000   0.62500000 
  0.50000000   0.75000000   0.75000000 
  0.50000000   0.75000000   0.87500000 
  0.50000000   0.87500000   0.00000000 
  0.50000000   0.87500000   0.12500000 
  0.50000000   0.87500000   0.25000000 
  0.50000000   0.87500000   0.37500000 
  0.50000000   0.87500000   0.50000000 
  0.50000000   0.87500000   0.62500000 
  0.50000000   0.87500000   0.75000000 
  0.50000000   0.87500000   0.87500000 
  0.62500000   0.00000000   0.00000000 
  0.62500000   0.00000000   0.12500000 
  0.62500000   0.00000000   0.25000000 
  0.62500000   0.00000000   0.37500000 
  0.62500000   0.00000000   0.50000000 
  0.62500000   0.00000000   0.62500000 
  0.62500000   0.00000000   0.75000000 
  0.62500000   0.00000000   0.87500000 
  0.62500000   0.12500000   0.00000000 
  0.62500000   0.12500000   0.12500000 
  0.62500000   0.12500000   0.25000000 
  0.62500000   0.12500000   0.37500000 
  0.62500000   0.12500000   0.50000000 
  0.62500000   0.12500000   0.62500000 
  0.62500000   0.12500000   0.75000000 
  0.62500000   0.12500000   0.87500000 
  0.62500000   0.25000000   0.00000000 
  0.62500000   0.25000000   0.12500000 
  0.62500000   0.25000000   0.25000000 
  0.62500000   0.25000000   0.37500000 
  0.62500000   0.25000000   0.50000000 
  0.62500000   0.25000000   0.62500000 
  0.62500000   0.25000000   0.75000000 
  0.62500000   0.25000000   0.87500000 
  0.62500000   0.37500000   0.00000000 
  0.62500000   0.37500000   0.12500000 
  0.62500000   0.37500000   0.25000000 
  0.62500000   0.37500000   0.37500000 
  0.62500000   0.37500000   0.50000000 
  0.62500000   0.37500000   0.62500000 
  0.62500000   0.37500000   0.75000000 
  0.62500000   0.37500000   0.87500000 
  0.62500000   0.50000000   0.00000000 
  0.62500000   0.50000000   0.12500000 
  0.62500000   0.50000000   0.25000000 
  0.62500000   0.50000000   0.37500000 
  0.62500000   0.50000000   0.50000000 
  0.62500000   0.50000000   0.62500000 
  0.62500000   0.50000000   0.75000000 
  0.62500000   0.50000000   0.87500000 
  0.62500000   0.62500000   0.00000000 
  0.62500000   0.62500000   0.12500000 
  0.62500000   0.62500000   0.25000000 
  0.62500000   0.62500000   0.37500000 
  0.62500000   0.62500000   0.50000000 
  0.62500000   0.62500000   0.62500000 
  0.62500000   0.62500000   0.75000000 
  0.62500000   0.62500000   0.87500000 
  0.62500000   0.75000000   0.00000000 
  0.62500000   0.75000000   0.12500000 
  0.62500000   0.75000000   0.25000000 
  0.62500000   0.75000000   0.37500000 
  0.62500000   0.75000000   0.50000000 
  0.62500000   0.75000000   0.62500000 
  0.62500000   0.75000000   0.75000000 
  0.62500000   0.75000000   0.87500000 
  0.62500000   0.87500000   0.00000000 
  0.62500000   0.87500000   0.12500000 
  0.62500000   0.87500000   0.25000000 
  0.62500000   0.87500000   0.37500000 
  0.62500000   0.87500000   0.50000000 
  0.62500000   0.87500000   0.62500000 
  0.62500000   0.87500000   0.75000000 
  0.62500000   0.87500000   0.87500000 
  0.75000000   0.00000000   0.00000000 
  0.75000000   0.00000000   0.12500000 
  0.75000000   0.00000000   0.25000000 
  0.75000000   0.00000000   0.37500000 
  0.75000000   0.00000000   0.50000000 
  0.75000000   0.00000000   0.62500000 
  0.75000000   0.00000000   0.75000000 
  0.75000000   0.00000000   0.87500000 
  0.75000000   0.12500000   0.00000000 
  0.75000000   0.12500000   0.12500000 
  0.75000000   0.12500000   0.25000000 
  0.75000000   0.12500000   0.37500000 
  0.75000000   0.12500000   0.50000000 
  0.75000000   0.12500000   0.62500000 
  0.75000000   0.12500000   0.75000000 
  0.75000000   0.12500000   0.87500000 
  0.75000000   0.25000000   0.00000000 
  0.75000000   0.25000000   0.12500000 
  0.75000000   0.25000000   0.25000000 
  0.75000000   0.25000000   0.37500000 
  0.75000000   0.25000000   0.50000000 
  0.75000000   0.25000000   0.62500000 
  0.75000000   0.25000000   0.75000000 
  0.75000000   0.25000000   0.87500000 
  0.75000000   0.37500000   0.00000000 
  0.75000000   0.37500000   0.12500000 
  0.75000000   0.37500000   0.25000000 
  0.75000000   0.37500000   0.37500000 
  0.75000000   0.37500000   0.50000000 
  0.75000000   0.37500000   0.62500000 
  0.75000000   0.37500000   0.75000000 
  0.75000000   0.37500000   0.87500000 
  0.75000000   0.50000000   0.00000000 
  0.75000000   0.50000000   0.12500000 
  0.75000000   0.50000000   0.25000000 
  0.75000000   0.50000000   0.37500000 
  0.75000000   0.50000000   0.50000000 
  0.75000000   0.50000000   0.62500000 
  0.75000000   0.50000000   0.75000000 
  0.75000000   0.50000000   0.87500000 
  0.75000000   0.62500000   0.00000000 
  0.75000000   0.62500000   0.12500000 
  0.75000000   0.62500000   0.25000000 
  0.75000000   0.62500000   0.37500000 
  0.75000000   0.62500000   0.50000000 
  0.75000000   0.62500000   0.62500000 
  0.75000000   0.62500000   0.75000000 
  0.75000000   0.62500000   0.87500000 
  0.75000000   0.75000000   0.00000000 
  0.75000000   0.75000000   0.12500000 
  0.75000000   0.75000000   0.25000000 
  0.75000000   0.75000000   0.37500000 
  0.75000000   0.75000000   0.50000000 
  0.75000000   0.75000000   0.62500000 
  0.75000000   0.75000000   0.75000000 
  0.75000000   0.75000000   0.87500000 
  0.75000000   0.87500000   0.00000000 
  0.75000000   0.87500000   0.12500000 
  0.75000000   0.87500000   0.25000000 
  0.75000000   0.87500000   0.37500000 
  0.75000000   0.87500000   0.50000000 
  0.75000000   0.87500000   0.62500000 
  0.75000000   0.87500000   0.75000000 
  0.75000000   0.87500000   0.87500000 
  0.87500000   0.00000000   0.00000000 
  0.87500000   0.00000000   0.12500000 
  0.87500000   0.00000000   0.25000000 
  0.87500000   0.00000000   0.37500000 
  0.87500000   0.00000000   0.50000000 
  0.87500000   0.00000000   0.62500000 
  0.87500000   0.00000000   0.75000000 
  0.87500000   0.00000000   0.87500000 
  0.87500000   0.12500000   0.00000000 
  0.87500000   0.12500000   0.12500000 
  0.87500000   0.12500000   0.25000000 
  0.87500000   0.12500000   0.37500000 
  0.87500000   0.12500000   0.50000000 
  0.87500000   0.12500000   0.62500000 
  0.87500000   0.12500000   0.75000000 
  0.87500000   0.12500000   0.87500000 
  0.87500000   0.25000000   0.00000000 
  0.87500000   0.25000000   0.12500000 
  0.87500000   0.25000000   0.25000000 
  0.87500000   0.25000000   0.37500000 
  0.87500000   0.25000000   0.50000000 
  0.87500000   0.25000000   0.62500000 
  0.87500000   0.25000000   0.75000000 
  0.87500000   0.25000000   0.87500000 
  0.87500000   0.37500000   0.00000000 
  0.87500000   0.37500000   0.12500000 
  0.87500000   0.37500000   0.25000000 
  0.87500000   0.37500000   0.37500000 
  0.87500000   0.37500000   0.50000000 
  0.87500000   0.37500000   0.62500000 
  0.87500000   0.37500000   0.75000000 
  0.87500000   0.37500000   0.87500000 
  0.87500000   0.50000000   0.00000000 
  0.87500000   0.50000000   0.12500000 
  0.87500000   0.50000000   0.25000000 
  0.87500000   0.50000000   0.37500000 
  0.87500000   0.50000000   0.50000000 
  0.87500000   0.50000000   0.62500000 
  0.87500000   0.50000000   0.75000000 
  0.87500000   0.50000000   0.87500000 
  0.87500000   0.62500000   0.00000000 
  0.87500000   0.62500000   0.12500000 
  0.87500000   0.62500000   0.25000000 
  0.87500000   0.62500000   0.37500000 
  0.87500000   0.62500000   0.50000000 
  0.87500000   0.62500000   0.62500000 
  0.87500000   0.62500000   0.75000000 
  0.87500000   0.62500000   0.87500000 
  0.87500000   0.75000000   0.00000000 
  0.87500000   0.75000000   0.12500000 
  0.87500000   0.75000000   0.25000000 
  0.87500000   0.75000000   0.37500000 
  0.87500000   0.75000000   0.50000000 
  0.87500000   0.75000000   0.62500000 
  0.87500000   0.75000000   0.75000000 
  0.87500000   0.75000000   0.87500000 
  0.87500000   0.87500000   0.00000000 
  0.87500000   0.87500000   0.12500000 
  0.87500000   0.87500000   0.25000000 
  0.87500000   0.87500000   0.37500000 
  0.87500000   0.87500000   0.50000000 
  0.87500000   0.87500000   0.62500000 
  0.87500000   0.87500000   0.75000000 
  0.87500000   0.87500000   0.87500000 
end kpoints
\end{lstlisting}
\section{Appendix: $\mathrm{HfP}_2$}
\subsection{TA.scf.in} 
\begin{lstlisting}
&CONTROL
calculation ='scf'
prefix = 'TA'
outdir = './bin'
pseudo_dir = './PP/'
verbosity='high'
/

&system
ibrav = 0
nat = 4
ntyp = 2
ecutwfc =55 !Ryberg
ecutrho =550
occupations = 'smearing'
smearing = 'mv'
degauss = 0.002

!SOC
noncolin = .TRUE.
lspinorb = .TRUE.
starting_magnetization(1) = 0
starting_magnetization(2) = 0


/

&ELECTRONS
conv_thr = 1.0d-7
mixing_beta = 0.495
diagonalization= 'david'
adaptive_thr=.true.
/

ATOMIC_SPECIES
Ta 180.94788 Ta_ONCV_PBE_fr.upf
As 74.9216 As_ONCV_PBE_fr.upf


ATOMIC_POSITIONS (crystal)
 Ta    0.25000   0.75000   0.50000 
 Ta    0.00000   0.00000   0.00000 
 As    0.66700   0.16700   0.33400 
 As    0.41700   0.41700   0.83400 


CELL_PARAMETERS (angstrom)
3.437000 -0.000000 -0.000000
-0.000000 3.437000 0.000000
-1.718500 -1.718500 5.828000


K_POINTS (automatic)
8 8 8 0 0 0 
\end{lstlisting}
\subsection{TA.nscf.in} 
\begin{lstlisting}
&CONTROL
calculation ='nscf'
prefix = 'TA'
outdir = './bin'
pseudo_dir = './PP/'
verbosity='high'
/

&system
ibrav = 0
nat = 4
ntyp = 2
ecutwfc =55 !Ryberg
ecutrho =550
!occupations = 'fixed'
occupations = 'smearing'
smearing = 'mv'
degauss = 0.002
nosym = .true.
nbnd = 120
!SOC
noncolin = .TRUE.
lspinorb = .TRUE.
starting_magnetization(1) = 0
starting_magnetization(2) = 0


/

&ELECTRONS
conv_thr = 1.0d-7
mixing_beta = 0.495
diagonalization= 'david'
adaptive_thr=.true.
/

ATOMIC_SPECIES
Ta 180.94788 Ta_ONCV_PBE_fr.upf
As 74.9216 As_ONCV_PBE_fr.upf



ATOMIC_POSITIONS (crystal)
 Ta    0.25000   0.75000   0.50000 
 Ta    0.00000   0.00000   0.00000 
 As    0.66700   0.16700   0.33400 
 As    0.41700   0.41700   0.83400 


CELL_PARAMETERS (angstrom)
3.437000 -0.000000 -0.000000
-0.000000 3.437000 0.000000
-1.718500 -1.718500 5.828000



K_POINTS (crystal)
64
  0.00000000   0.00000000   0.00000000 1
  0.00000000   0.00000000   0.25000000 1
  0.00000000   0.00000000   0.50000000 1
  0.00000000   0.00000000   0.75000000 1
  0.00000000   0.25000000   0.00000000 1
  0.00000000   0.25000000   0.25000000 1
  0.00000000   0.25000000   0.50000000 1
  0.00000000   0.25000000   0.75000000 1
  0.00000000   0.50000000   0.00000000 1
  0.00000000   0.50000000   0.25000000 1
  0.00000000   0.50000000   0.50000000 1
  0.00000000   0.50000000   0.75000000 1
  0.00000000   0.75000000   0.00000000 1
  0.00000000   0.75000000   0.25000000 1
  0.00000000   0.75000000   0.50000000 1
  0.00000000   0.75000000   0.75000000 1
  0.25000000   0.00000000   0.00000000 1
  0.25000000   0.00000000   0.25000000 1
  0.25000000   0.00000000   0.50000000 1
  0.25000000   0.00000000   0.75000000 1
  0.25000000   0.25000000   0.00000000 1
  0.25000000   0.25000000   0.25000000 1
  0.25000000   0.25000000   0.50000000 1
  0.25000000   0.25000000   0.75000000 1
  0.25000000   0.50000000   0.00000000 1
  0.25000000   0.50000000   0.25000000 1
  0.25000000   0.50000000   0.50000000 1
  0.25000000   0.50000000   0.75000000 1
  0.25000000   0.75000000   0.00000000 1
  0.25000000   0.75000000   0.25000000 1
  0.25000000   0.75000000   0.50000000 1
  0.25000000   0.75000000   0.75000000 1
  0.50000000   0.00000000   0.00000000 1
  0.50000000   0.00000000   0.25000000 1
  0.50000000   0.00000000   0.50000000 1
  0.50000000   0.00000000   0.75000000 1
  0.50000000   0.25000000   0.00000000 1
  0.50000000   0.25000000   0.25000000 1
  0.50000000   0.25000000   0.50000000 1
  0.50000000   0.25000000   0.75000000 1
  0.50000000   0.50000000   0.00000000 1
  0.50000000   0.50000000   0.25000000 1
  0.50000000   0.50000000   0.50000000 1
  0.50000000   0.50000000   0.75000000 1
  0.50000000   0.75000000   0.00000000 1
  0.50000000   0.75000000   0.25000000 1
  0.50000000   0.75000000   0.50000000 1
  0.50000000   0.75000000   0.75000000 1
  0.75000000   0.00000000   0.00000000 1
  0.75000000   0.00000000   0.25000000 1
  0.75000000   0.00000000   0.50000000 1
  0.75000000   0.00000000   0.75000000 1
  0.75000000   0.25000000   0.00000000 1
  0.75000000   0.25000000   0.25000000 1
  0.75000000   0.25000000   0.50000000 1
  0.75000000   0.25000000   0.75000000 1
  0.75000000   0.50000000   0.00000000 1
  0.75000000   0.50000000   0.25000000 1
  0.75000000   0.50000000   0.50000000 1
  0.75000000   0.50000000   0.75000000 1
  0.75000000   0.75000000   0.00000000 1
  0.75000000   0.75000000   0.25000000 1
  0.75000000   0.75000000   0.50000000 1
  0.75000000   0.75000000   0.75000000 1

\end{lstlisting}
\subsection{TA.win} 
\begin{lstlisting}
write_hr = .TRUE.
write_xyz = .TRUE.
!wannier_plot = .TRUE. 
spinors = .TRUE.
num_wann = 120
dis_num_iter=1000
!trial_step=50
num_iter = 200
!guiding_centres = .TRUE.

!exclude_bands: 41-120


begin unit_cell_cart
3.437000 -0.000000 -0.000000
-0.000000 3.437000 0.000000
-1.718500 -1.718500 5.828000
end unit_cell_cart


begin atoms_frac
 Ta    0.25000   0.75000   0.50000 
 Ta    0.00000   0.00000   0.00000 
 As    0.66700   0.16700   0.33400 
 As    0.41700   0.41700   0.83400 
end atoms_frac


begin projections 
random 
end projections


mp_grid = 4 4 4 
begin kpoints
  0.00000000   0.00000000   0.00000000 
  0.00000000   0.00000000   0.25000000 
  0.00000000   0.00000000   0.50000000 
  0.00000000   0.00000000   0.75000000 
  0.00000000   0.25000000   0.00000000 
  0.00000000   0.25000000   0.25000000 
  0.00000000   0.25000000   0.50000000 
  0.00000000   0.25000000   0.75000000 
  0.00000000   0.50000000   0.00000000 
  0.00000000   0.50000000   0.25000000 
  0.00000000   0.50000000   0.50000000 
  0.00000000   0.50000000   0.75000000 
  0.00000000   0.75000000   0.00000000 
  0.00000000   0.75000000   0.25000000 
  0.00000000   0.75000000   0.50000000 
  0.00000000   0.75000000   0.75000000 
  0.25000000   0.00000000   0.00000000 
  0.25000000   0.00000000   0.25000000 
  0.25000000   0.00000000   0.50000000 
  0.25000000   0.00000000   0.75000000 
  0.25000000   0.25000000   0.00000000 
  0.25000000   0.25000000   0.25000000 
  0.25000000   0.25000000   0.50000000 
  0.25000000   0.25000000   0.75000000 
  0.25000000   0.50000000   0.00000000 
  0.25000000   0.50000000   0.25000000 
  0.25000000   0.50000000   0.50000000 
  0.25000000   0.50000000   0.75000000 
  0.25000000   0.75000000   0.00000000 
  0.25000000   0.75000000   0.25000000 
  0.25000000   0.75000000   0.50000000 
  0.25000000   0.75000000   0.75000000 
  0.50000000   0.00000000   0.00000000 
  0.50000000   0.00000000   0.25000000 
  0.50000000   0.00000000   0.50000000 
  0.50000000   0.00000000   0.75000000 
  0.50000000   0.25000000   0.00000000 
  0.50000000   0.25000000   0.25000000 
  0.50000000   0.25000000   0.50000000 
  0.50000000   0.25000000   0.75000000 
  0.50000000   0.50000000   0.00000000 
  0.50000000   0.50000000   0.25000000 
  0.50000000   0.50000000   0.50000000 
  0.50000000   0.50000000   0.75000000 
  0.50000000   0.75000000   0.00000000 
  0.50000000   0.75000000   0.25000000 
  0.50000000   0.75000000   0.50000000 
  0.50000000   0.75000000   0.75000000 
  0.75000000   0.00000000   0.00000000 
  0.75000000   0.00000000   0.25000000 
  0.75000000   0.00000000   0.50000000 
  0.75000000   0.00000000   0.75000000 
  0.75000000   0.25000000   0.00000000 
  0.75000000   0.25000000   0.25000000 
  0.75000000   0.25000000   0.50000000 
  0.75000000   0.25000000   0.75000000 
  0.75000000   0.50000000   0.00000000 
  0.75000000   0.50000000   0.25000000 
  0.75000000   0.50000000   0.50000000 
  0.75000000   0.50000000   0.75000000 
  0.75000000   0.75000000   0.00000000 
  0.75000000   0.75000000   0.25000000 
  0.75000000   0.75000000   0.50000000 
  0.75000000   0.75000000   0.75000000 
end kpoints
\end{lstlisting}

\end{document}